\documentclass[12pt]{article}
\include{epsf}
\def\beq{\begin{eqnarray}}
\def\eeq{\end{eqnarray}}
\usepackage{amssymb}
\usepackage{epsfig}

\begin{document}

\title{A one-dimensional toy model of globular clusters}

\author{D.~Fanelli$^1$\thanks{e-mail: fanelli@nada.kth.se}, 
M.~Merafina$^2$\thanks{e-mail: Marco.Merafina@roma1.infn.it},
S.~Ruffo$^3$\thanks{e-mail: ruffo@avanzi.de.unifi.it}}
\maketitle 

\begin{center}
\begin{tabular}{ll}
$^1$ & Department of Numerical Analysis and Computer Science, KTH,\\ 
& $\qquad$ S-100 44 Stockholm, Sweden\\
$^2$ & Dipartimento di Fisica, Universit\'a di Roma ``La Sapienza",\\
& $\qquad$ Piazzale Aldo Moro, 2 - I-00185 Roma, Italy\\
$^3$ & Dipartimento di Energetica ``Sergio Stecco", Universit\'a \\
& $\qquad$ di Firenze, INFM and INFN, Via S. Marta, 3 - I-50139 Firenze, Italy
\end{tabular}
\end{center}
\date{}

\begin{abstract} 
We introduce a one-dimensional toy model of globular clusters. The
model is a version of the well-known gravitational sheets system, where
we take additionally into account mass and energy loss by evaporation of stars
at the boundaries. Numerical integration by the ``exact" event-driven 
dynamics is performed, for initial uniform density and Gaussian random 
velocities. Two distinct quasi-stationary asymptotic regimes are 
attained, depending on the initial energy of the system. 
We guess the forms of the density and velocity profiles which fit 
numerical data extremely well and allow to perform an independent 
calculation of the self-consistent gravitational potential. Some power-laws 
for the asymptotic number of stars and for the collision times are suggested.
\end{abstract}


\section{Introduction}
\label{s:introduction}

Globular clusters are gravitationally bound concentrations of large numbers of 
stars, spherically distributed in space. They orbit around a galaxy 
spending most of the time in the galactic halo \cite{Chandra}.
The most important elements governing globular clusters structure are 
two-body relaxation and  truncation due to tidal forces.
Different dynamical models considering these specific 
phenomena, have been investigated both analytically~\cite{Chandra,henon,Inagaki,King} and 
numerically~\cite{Cohn,Inagaki1,Miller1}. 
For an extended discussion, see~\cite{meylan} and Refs. therein.

Dynamical evolution  causes stars to escape as an effect of the 
gravitational interaction with the nearby galaxies. This evaporation process
drives the cluster towards a  configuration with a high density core and the 
velocity dispersion of stars in the bulk can increase  without limit. 
This phenomenon is known as {\it gravothermal catastrophe} and its study 
goes back to Antonov~\cite{Antonov} and to Lynden-Bell and Wood~\cite{Lynd}.

Referring to the pioneering work of Chandrasekhar
\cite{Chandra}, it is possible to calculate the perturbations
induced by stellar encounters on star motion. 
This is done by means of a diffusion model, which allows to derive a quantitative
description of changes of star velocities in terms of single encounters.
Considering  weak encounters, i.e. solving the diffusion model in the Fokker-Planck 
approximation, King~\cite{King} found the following expression for the velocity 
profile
\begin{equation}
\label{eq:df}
\left\{
\begin{array}{ll}
f_0 (v) = A\left(\exp\left(\frac{-v^2}{2\sigma^2}\right)-
\exp\left(\frac{-v_c^2}{2\sigma^2}\right) \right) & \mbox{for $v\le v_c$} \\
\\
f_0  = 0 & \mbox{for $v>v_c$,}
\end{array} 
\right \}
\end{equation}
\noindent
where $v_c$ is a cutoff velocity of the stars, $\sigma$ is
the one-dimensional velocity dispersion and $A$ a 
normalization constant. King's models have been shown 
to be in agreement with  the observed brightness surface
profiles of globular clusters~\cite{King}. 
Further developments of King's model~\cite{meylan} consider the same functional
form (truncated Gaussian) applied to the gravitational energy, and hence propose
a general form for the distribution function $f({\bf x},{\bf v})$. We
will not consider here these extensions.

In this paper we discuss a simplified one-dimensional $N$-body model
which reproduces King's distribution. We let the particles, all
of equal mass $m$, interact through the one-dimensional gravitational
potential $V =2 \pi G m^2 |x|$, where $G$ is the gravitational constant. Bearing 
in mind the comparison with globular clusters, we
imitate the effect of galactic tidal forces by introducing a finite 
cut-off in positions. Thus, the evaporation of stars from the system is 
the only ``dissipative" effect we consider.  This is 
enough to drive the system towards an asymptotic {\it non stationary} regime, 
which we analyze in detail, and which reveals striking similarity with 
King's model.

The main difference between the simplified one-dimensional model considered 
here and a more realistic three-dimensional one, is the lack of any 
singularity of the potential at the origin. The presence of a finite lower bound for the
$1D$ potential makes less energy available to support the evaporation process
as the system cools down. In the $3D$ case an infinite amount of energy can 
indeed be extracted from the singular pair-wise gravitational interaction,
which is the main origin of the gravothermal catastrophe. 
This is the reason why the model we discuss cannot reproduce the 
core collapse corresponding to the gravothermal catastrophe.

In the next Section we present the model and the results concerning
velocity distribution and density profiles. 
Section~\ref{Scal} is devoted to the discussion of power-laws for the number 
of particles in the cluster. Finally, in Section~\ref{s:concl} we draw
some conclusions.

\section{One-dimensional model}
\label{model}

Let us consider a one-dimensional classical Newtonian self-gravitating  
system of $N$ particles with equal mass $m$, with Hamiltonian~\cite{hohl}
\begin{equation}
H= \frac{1}{2} m \sum_i^N  v^2_i + 2 \pi G m^2 \sum_{j<i} |x_i - x_j|~,
\label{Ham}
\end{equation}
where $x_i$ is the position and $v_i$ the velocity of $i$-th star. $G$ is the 
universal gravitational constant. We choose in the following $m=1$ and $2 \pi G=1$.
This system has been recently the subject of intensive investigations~\cite{self}.
Particle accelerations are constant inbetween two collisions 
and are proportional to the net difference of particles respectively on the 
right and on the left. When a collision occurs particles cross each other,
or, equivalently, collisions are elastic.
This particle approach is known to corresponds in the continuum limit 
($N \to \infty$) to the Vlasov-Poisson equations for the distribution 
function $f(x,v)$
\begin{eqnarray}
&& \frac{\partial f}{\partial t} + v \frac{\partial f}{\partial x} 
- \frac{\partial V}{\partial x} \frac{\partial f}{\partial v}=0 \nonumber \\
&& \frac{\partial^2 V}{\partial x^2}= 4 \pi G m \int f(x,v) dv~,
\label{Vlasov}
\end{eqnarray}
where $V(x)$ is the self-consistent gravitational potential.

We add the following features to the model
\begin{itemize}
\item Particles are confined in a box of size $L$, {i.e.} 
$x_i \in [-L/2,L/2]$
\item The effect of tidal forces induced by the parent galaxy 
is imitated by requiring that each time a 
particle reaches the boundary of the box with a finite velocity, 
it drops out of the system, which therefore experiences a mass 
and energy loss (``evaporation"). This last feature implies that
system (\ref{Vlasov}) is solved with absorbing boundary conditions.
\end{itemize} 
The numerical implementation is based on an ``event-driven''
scheme, first introduced in plasma physics~\cite{Feix}, which is adapted 
to the present case as follows. The algorithm looks for the particles 
which collide the first and for the time when the event occurs, $t_{coll}$. 
Then it computes the first ``evaporation time", $t_{evap}$, and makes 
the system evolve until the minimum, $t_{min}$, between the collision and
the evaporation time is reached.
Once the system experiences evaporation, the total mass is reduced and 
the escaping particle stops interacting with the residual bulk.
By rescaling the position and the velocity of each particle, {\it i.e.} 
introducing a local dissipation, we maintain the position of the center of 
mass fixed and its velocity to zero. This means we simply translate velocity 
and position of the remaining particles to keep the system centered in
position and momentum space. A particle can escape from the system as  
a result of this rescaling: this possibility is taken into account even if 
it is has a low probability.

Evaporation is a singular event, which, in fact, marks the transition between
two self-gravitating system having a different number of particles and
energy. We remark that the integration scheme is ``exact".
Time $t$ elapsed from the initial configuration is obtained by summing all 
values of $t_{min}$ up to the last event.   

In all numerical experiments, $N(0)$ particles are initially uniformly 
distributed in the box and the initial velocities are Gaussian i.i.d random 
variables, with the temperature $T_0$ given by twice the average kinetic 
energy.

\begin{figure}
\begin{center}
\psfig{figure=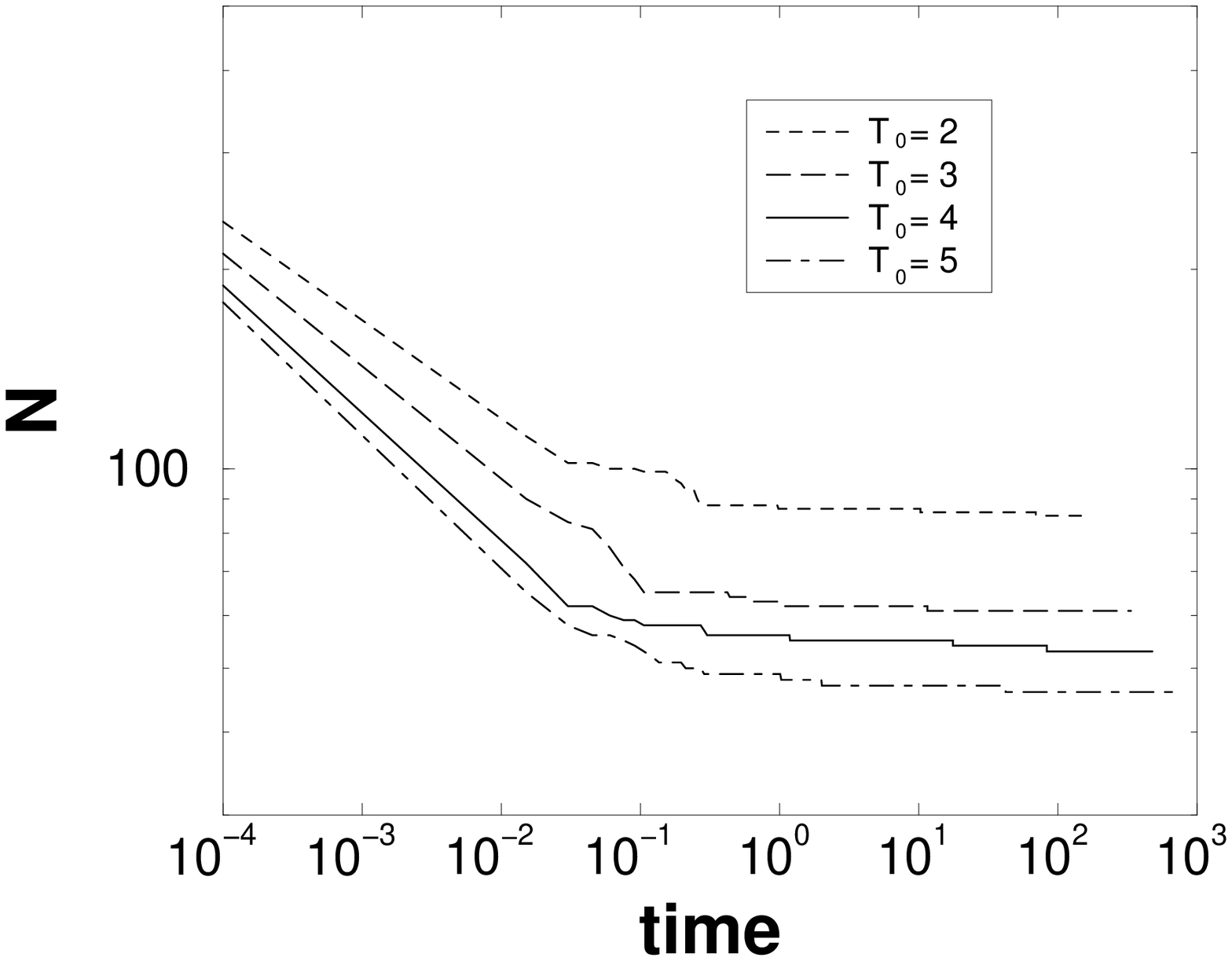,height=8truecm,width=10truecm}
\caption{\label{logN-logt} Plots of $N$ vs. $time$ for increasing initial
temperatures, with $N(0)=400$. Temperature and time are expressed in
arbitrary units.}
\end{center}
\end{figure}

In Fig.~\ref{logN-logt} we show the time evolution of the number of
particles $N(t)$ which remain inside the box up to time $t$. After an
abrupt decrease of $N(t)$, which strongly depends on the initial
condition, the system reaches a state where rare evaporations
are present, making $N(t)$  decrease much slowlier. 
Our numerical experiments show that such a {\it quasi-stationary}
state lives indefinitely, although we cannot exclude that, finally,
$N(t)$ relaxes to an asymptotic value $N_{as}$. As the best
approximation for this value, we take the one the system reaches in the
longest computer runs.

\begin{figure}
\begin{center}
\psfig{figure=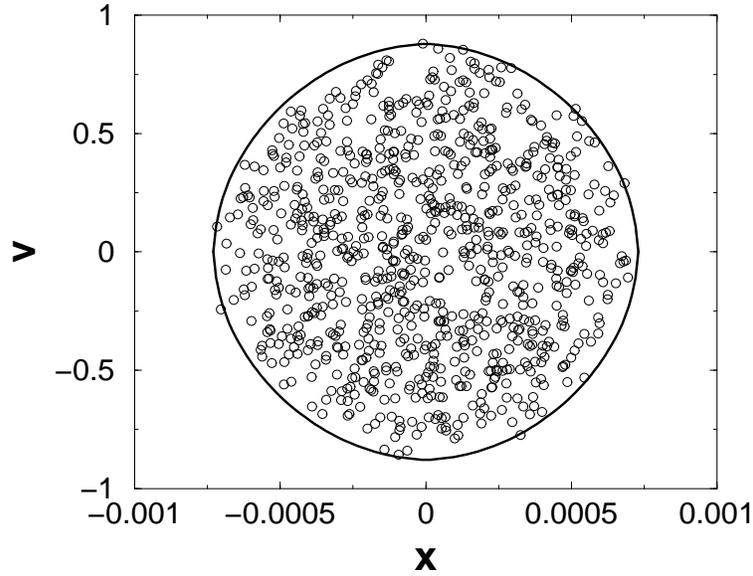,height=8truecm,width=10truecm} 
\caption{\label{fasi} 
Phase-space plot for a system of $N(0)=1500$ particles after 
$30 \times 10^6$ collisions. Here $T_0=0.4$, $L=0.0015$ and 
$N_{as}=886$.
The quantities $x$ and $v$ are expressed in arbitrary units.
The full line which contains all the points is the theoretical
prediction (\ref{contour})}
\end{center}
\end{figure}

Given $L$, for large enough $T_0$ the system approaches a state  
characterized by a single cluster, which adapts itself to the size of the 
box. In Fig.~\ref{fasi} we show the phase-space portrait for a system in 
its late stage evolution, when the most energetic particles 
have dropped out from the box and the system has relaxed to an asymptotic 
{\it plateau}, as the ones reported in Fig.~\ref{logN-logt}.
The particles are almost uniformly distributed within
a bounded region of the phase plane. Both this fact and the shape of the 
contour suggest a possible connection with the so called water-bag (WB) 
distribution~\cite{hohl,dep}. 
This is a {\it stationary} solution of the Vlasov-Poisson 
system (\ref{Vlasov}), $f^{WB}(x,v)$, which is constant in a simply connected domain
$\Omega$ of the phase-plane and strictly zero outside.

Adopting the notation of Ref.~\cite{hohl}, we call 
respectively $x_s$ and $v_s$ the maximum position and velocity of the WB.  
The potential $V(x)$ for such a distribution is then implicitly 
specified by the following integral equation
\begin{equation}
\label{potWB}
x=\frac{\epsilon^{\frac{3}{4} }}{N_{as}} \int^{V(x)}_{0} 
\left[\epsilon^{ \frac{3}{2} } - (\epsilon-\zeta)^{ \frac{3}{2} } \right]^
{ -\frac{1}{2} } d \zeta~.
\end{equation}
The maximal energy of the water-bag, $\epsilon$, is such that, if
we express $f(x,v)$ in terms of the energy $u=v^2/2+V(x)$, $f(x,v)=F(u)=0$
if $u>\epsilon$. The energy $\epsilon$ is related to $x_s$ and $v_s$
by $\epsilon=v_s^2/2$, $V(x_s)=\epsilon $ and the zero energy level
is fixed by requiring that $V(0)=0$. 
The density profile, $\rho^{WB}(x)$, is expressed as 
function of the potential
\begin{equation}
\label{densWB}
\rho^{WB}(x) = \frac{3 N_{as}^2}{8 \epsilon^{3/2}} (\epsilon - V)^{1/2}~.
\end{equation}
Since the distribution function $f(x,v)$ is constant over $\Omega$, it follows 
immediately
\begin{equation}
\rho^{WB}(x)=\int^{v^{+}(x)}_{-v^{+}(x)} f(x,v) dv = 2 c v^{+}(x)
\end{equation}
where $v^{+}(x)$ represents the profile of the upper branch of the WB contour,
which we assumed to be symmetric, and $c=3N_{as}^2/(16 \sqrt{2} \epsilon^{3/2})$. 
Using Eq.~(\ref{densWB}), this implies
\begin{equation} 
\label{contour}
v^{+}(x) = \sqrt{2(\epsilon - V)}~. 
\end{equation}
To compute the velocity contour $v^{+}(x)$, we need to know $V(x)$, which we
do by solving Eq.~(\ref{potWB}) by an adaptive 
recursive Newton-Cotes 8 panel rule with tolerance $10^{-7}$.
This velocity contour is drawn in Fig.~\ref{fasi} and, as predicted, encloses
all phase points. In deriving $V(x)$ we have taken the value
$v_s$ from the cluster phase plot; this is the only phenomenological
input in this calculation and the agreement with the data has to be 
considered quite satisfactory.

\begin{figure}
\begin{center}
\psfig{figure=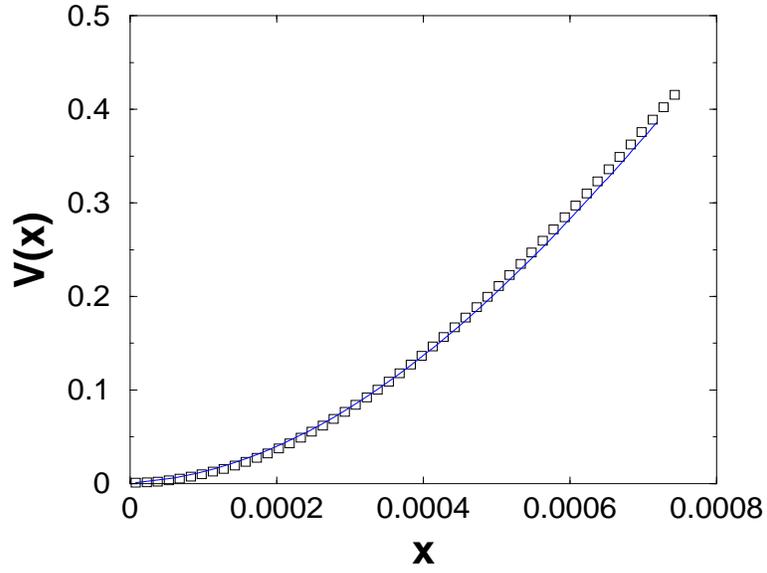, height=8truecm,width=10truecm} 
\caption{\label{WB} Gravitational potential 
calculated numerically using Eq.~(\ref{potenznumeric}) 
(squares) and analytically by solving the integral 
equation (\ref{potWB}) with $v_s=0.88$ (full line).
Only the region of positive $x$ is drawn, in arbitrary units.}  
\end{center}
\end{figure}

Moreover, we can compare the theoretically derived potential $V(x)$
with the one computed directly from the asymptotic positions
\begin{equation} 
V ( x_i ) = \sum_j | x_i - x_j |~.
\label{potenznumeric}
\end{equation}
We need, of course, to perform a vertical shift 
to fix the zero in the origin. The result of formula (\ref{potenznumeric}) is 
reported in Fig.~\ref{WB} together with the theoretically
derived potential. The agreement is really good. We are thus led to
conclude that our asymptotic state is well described by a water-bag.
However, this latter is a stationary solution of the self-gravitating
1D system, while in our simulations we continue to observe some
particle evaporations even at very long times. This is why
we have called our asymptotic state {\it quasi-stationary} and its
description in terms of a water-bag distribution can only be
approximate.

\begin{figure}
\begin{center}
\psfig{figure=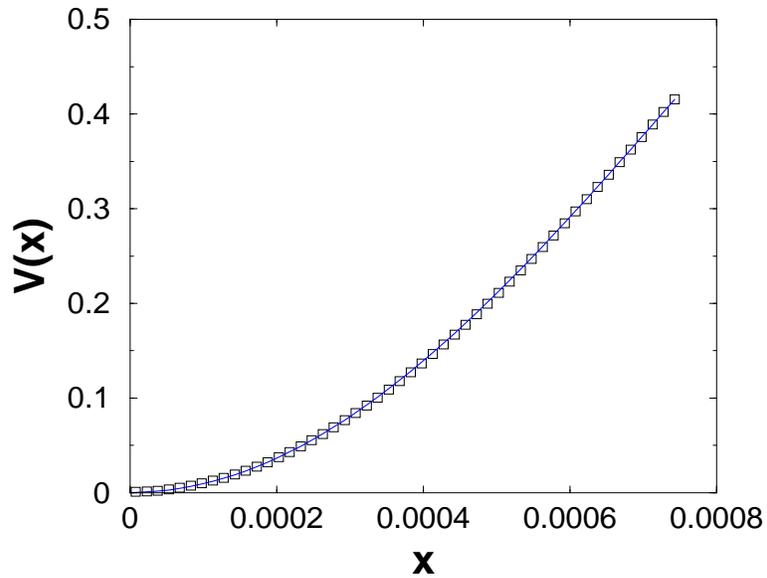, height=8truecm,width=10truecm}
\caption{\label{potenzialefit} Gravitational potential 
calculated numerically using Eq.~(\ref{potenznumeric}) 
(squares) and by a numerical fit which uses Eq.~(\ref{potenzprim})
(full line), where $B=731.2$ is the only free parameter.}  
\end{center}
\end{figure}

\begin{figure}
\begin{center}
\psfig{figure=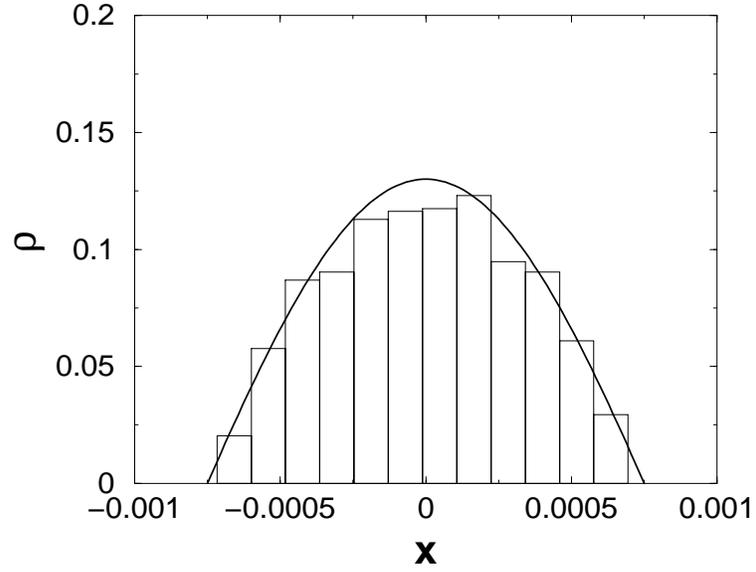, height=8truecm,width=10truecm }
\caption{\label{density} Normalized histogram of positions as 
derived from the phase-space plot in Fig.~\ref{fasi}. The solid line is
Eq.~(\ref{dens}), with $B=731.2$ as previously.
Position $x$ is expressed in arbitrary units.} 
\end{center}
\end{figure}

\begin{figure}
\begin{center}
\psfig{figure=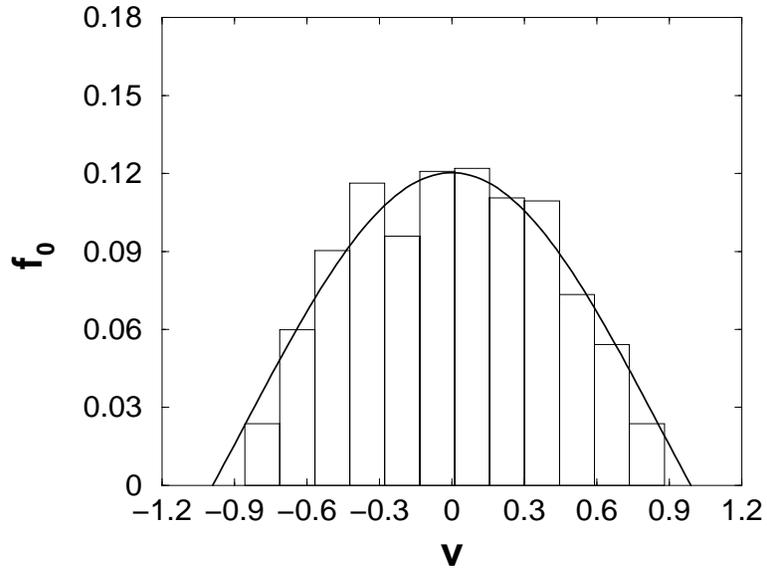, height=8truecm,width=10truecm}
\caption{\label{veloc} Normalized histogram of velocities as 
derived from the phase-space plot in Fig.~\ref{fasi}.
The solid line is the fit obtained using Eq.~(\ref{eq:df}) 
with $\sigma = 0.87, v_c = 0.98$.
Velocity $v$ is expressed in arbitrary units.}
\end{center}
\end{figure}

An alternative treatment of the asymptotic state is based on
King's formula (\ref{eq:df}). In this case one does not
try to reproduce the full distribution function, but just
its projections along the $x$ and $v$ axis: $\rho(x)=\int f dv$
and $f_0(v)=\int f dx$.
Following the standard derivation of the equilibrium 
isothermal distribution~\cite{ryb}, we are led 
to introduce an analytical ans\"atz for the density profile 
\begin{equation}
\label{dens}
\left\{
\begin{array}{ll}
\rho(x) = A ({cosh^{-2}(Bx)}-{cosh^{-2}(B \frac{L}{2})}) & \mbox{for
$|x|<\frac{L}{2}$} \\
\\
\\
\rho(x) = 0 & \mbox{for $|x|>\frac{L}{2}$~,}
\end{array}
\right\}
\end{equation}
where the normalization $A$ is fixed by $\int_{-L/2}^{L/2} \rho(x) dx = N_{as}$. 
For the velocities we take King's distribution, specified in Eq.~(\ref{eq:df}).
Assuming $\rho(x)$ as in Eq.~(\ref{dens}) we can derive a close analytical 
expression for the potential. 
For a  one-dimensional system, the following relation holds in the 
continuum limit
\begin{equation}
 V ( x ) = \int^{\frac{L}{2}}_{ - \frac{L}{2} } |y - x| \rho ( y ) dy~.
\label{integpotenz}
\end{equation}
Inserting Eq.~(\ref{dens}) into Eq.~(\ref{integpotenz}) and performing
the integral, we get
\begin{equation}
V ( x ) = 
\\
          V_0\left[\frac{L}{B} \tanh ( B \frac{L}{2} ) -
          ( x^2 +  \frac{L^2}{4} ) \cosh ^ {-2} (B \frac{L}{2}) + \\
\\
           \frac{2}{B^2} \ln (\frac{\cosh (B x) }{\cosh (B  \frac{L}{2} )})\right]~, 
\label{potenzprim}
\end{equation}
with
\begin{equation}
V_0=\frac{N}{\frac{2}{B} \tanh(B \frac{L}{2}) - L \cosh^{-2}( B \frac{L}{2} ) }~.
\end{equation}
$V(x)$ is quadratic for small $x$. 
To verify the reliability of our guess we re-analyze the data
previously discussed in connection with the water-bag distribution.
In Fig.~\ref{potenzialefit} we plot the potential calculated numerically 
from Eq.~(\ref{potenznumeric}) together with a one-parameter fit, 
using Eq.~(\ref{potenzprim}). Again the agreement with the data
is very good and even apparently superior to the one obtained using
the WB picture. This is simply due to the fact that here we perform a one 
parameter fit, while, in the previous discussion, $v_s$ was arbitrarily 
deduced from the phase space analysis.
As a cross-check, we introduce in Eq.~(\ref{dens}) the coefficient $B$ 
determined from the fit of the potential. The resulting density profile is 
plotted in Fig.~\ref{density} and it agrees with the 
normalized histogram of particles position.
Finally, an histogram of the velocity is represented in Fig.~\ref{veloc}.
The reverse-cup shape due to the cut-off of the tails is evident. The solid 
line in Fig.~\ref{veloc} is a numerical fit which uses the expression 
of Eq.~(\ref{eq:df}) with $v_c$ and $\sigma$ as free parameters.

\begin{figure}
\begin{center}
\psfig{figure=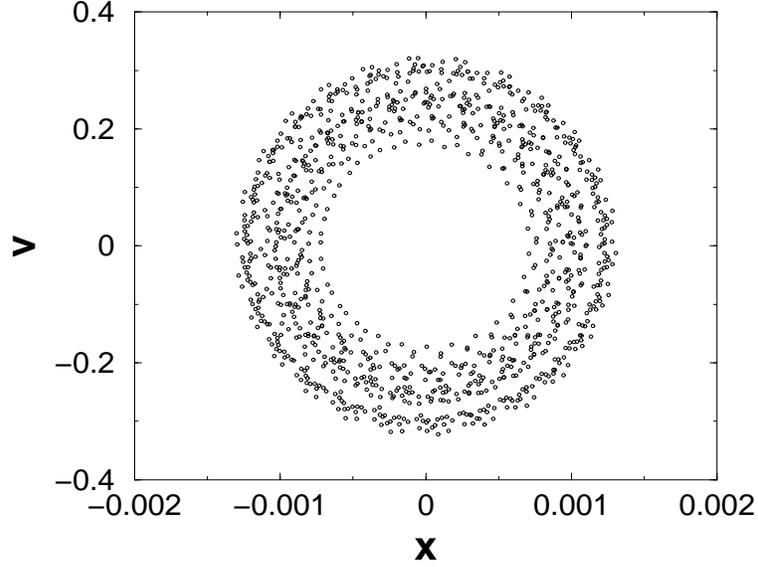,height=8truecm,width=10truecm}
\caption{\label{xv} Asymptotic orbit of a single particle for 
$N(0)=600$, $L=0.0035$. Positions and velocities are 
expressed in arbitrary units.}
\end{center}
\end{figure}

As a side remark we observe that, coherently with the observed form of 
the potential, each particle oscillates
almost harmonically inside the box. This can be seen by looking at the 
asymptotic orbit of a single particle (Fig.~\ref{xv}).
The slight diffusion of the orbit is the signature of the interaction
with the other particles, which induces a weak chaoticity.

\begin{figure}
\begin{center}
\psfig{figure=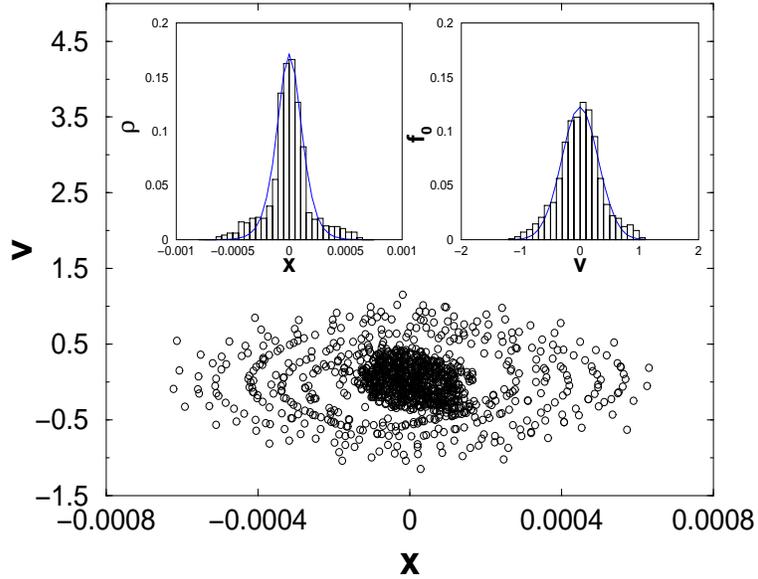,height=8truecm,width=10truecm} 
\caption{\label{ryb} Phase-space plot for a system of $N(0)=1500$ 
particles after $30 \times 10^6$ collisions. Here $T_0=0.02$, $L=0.0015$ 
and $N_{as}=1167$.  
Left insert: normalized histogram of positions. The solid line is a 
fit which uses Eq.~(\ref{dens}) where $B=6845.7$.
Right insert: normalized histogram of velocities. 
The solid line is the fit obtained using Eq.~(\ref{eq:df}) 
with $\sigma = 0.3$, $v_c = 1.9$.
All quantities are expressed in arbitrary units.  }
\end{center}
\end{figure}

A further aspect that we have tested is the dependence of the dynamics on the 
initial temperature, $T_0$. Indeed, for small values of $T_0$, the system 
shows a pronounced collapse, which leads to a massive central core, 
as it is clearly displayed in the main plot of Fig.~\ref{ryb}.
This phase-space distribution significantly differs 
from the one in Fig.~\ref{fasi} and cannot be represented by a water-bag.
The histograms of positions and velocities are computed and plotted 
in the right and left inserts, respectively.
Both the density and the velocity profiles are very well reproduced by 
a numerical fit based on our ans\"atz (\ref{dens}) and 
on King's distribution (\ref{eq:df}).

In conclusion, the ans\"atz we have introduced shows a good agreement with 
numerical data for all values of $T_0$ we have simulated, while the 
water-bag distribution
fails to reproduce the velocity and density profiles at very 
low initial temperature. However, in the high temperature range, the
water-bag treatment is superior, because it leads to an accurate
description of the full distribution $f(x,v)$.

\section{Scaling laws}
\label{Scal}

In this Section we discuss some numerically found scaling laws which
do not have presently a theoretical justification but which are an important
signature of the presence of a finite box.
   
We follow the system until it reaches the asymptotic state. Being
$N_{coll}$ the number of collisions, we define  the 
``average collision time'', $\tau = \frac{t}{N_{coll}}$. 

\begin{figure}
\begin{center}
\psfig{figure=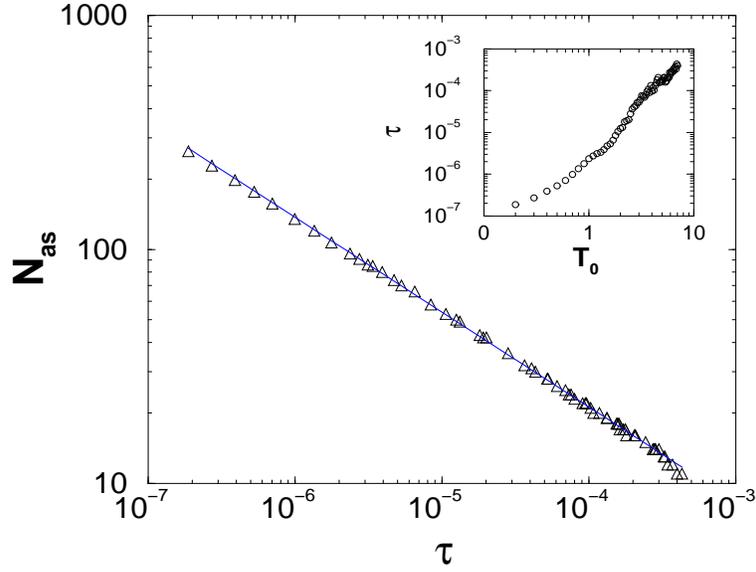,height=8truecm,width=10truecm}
\caption{\label{Nt} $N_{as}$ vs. $\tau$ in log-log scale. Here 
$\tau$ is defined as the ratio $\frac{t}{N_{coll}}$. Each point
refers to a different $T_0$ while $N_{coll}$ is fixed. The solid line 
represents a power-law fit with the  slope $a=-0.4$. In the upper 
right corner insert $\tau$ vs. $T_0$ is represented in a log-log scale. 
The quantities $T_0$ and $\tau$ are expressed in arbitrary units.}
\end{center}
\end{figure}

In the main plot of Fig~\ref{Nt}  we represent $N_{as}$ as a function 
of $\tau$. Each point refers to a different value of the initial 
temperature $T_0$, varying form $0.2$ to $7$, while $N_{coll}$ 
is maintained constant for each realization.  
The initial temperature controls the rate of evaporation at a very 
early stage of the evolution. 
Larger values of $T_0$, produce higher mass loss, inducing the system 
to relax to a quasi-stationary state characterized by less residual 
particles $N_{as}$ (see Fig. \ref{logN-logt}). 
This process has consequences at the dynamical level, determining a 
larger mean free path  and consequently a larger value of $\tau$. 
The curve in Fig~\ref{Nt} is consistent with this qualitative picture, 
showing a power-law decay with exponents $a=-0.4$ which is valid 
over more than two decades. In the insert of Fig.~\ref{Nt}, 
we plot $\tau$ vs. $T_0$.

\begin{figure}
\begin{center}
\psfig{figure=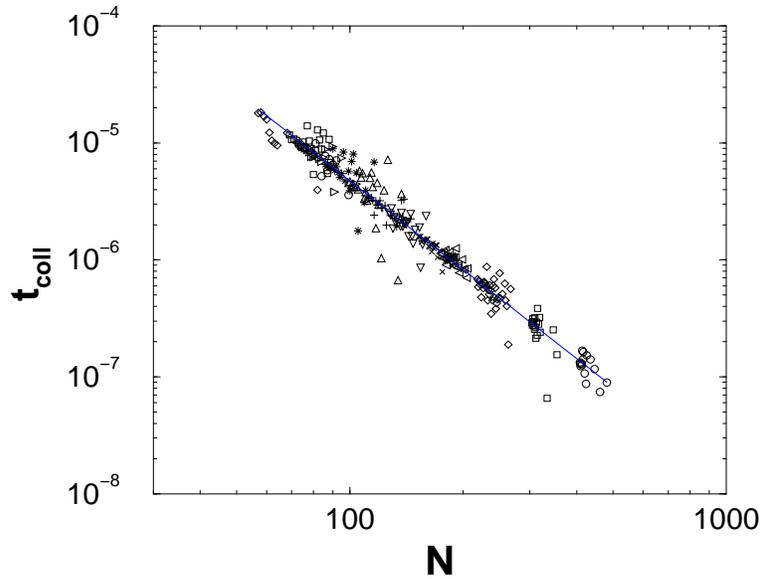,height=8truecm,width=10truecm}
\caption{\label{tcollN} $t_{coll}$ vs. $N$ in log-log scale for 
$N(0)=600$ $L=0.0035$. Different symbols
refer to different initial values of the temperature $T_0$.
$t_{coll}$ is expressed in arbitrary units.}
\end{center}
\end{figure}

We have also checked the dependence on $N$ of the collision time $t_{coll}$, 
defined as an average of all the collision times corresponding to each
fixed value of $N$, {\it i.e.} inbetween two successive evaporations. 
In Fig.~\ref{tcollN} we plot the results of different numerical 
experiments where we vary the initial temperature $T_0$. 
The dependence of $t_{coll}$ on $N$ is again a power-law with exponent $b=-2.5$.
Note that $b \sim 1/a$, hence Fig.~\ref{Nt} can be thought as a 
macroscopic averaged image of the microscopic properties 
shown in Fig.~\ref{tcollN}.

\section{Conclusions}
\label{s:concl}

We have introduced a one-dimensional toy model of globular
clusters with an emphasis on the evaporation process. 
With this in mind, we  have discussed the effect of introducing a finite
size box in a classical one dimensional self-gravitating medium.
The dynamics of the system has been investigated for a special class of 
initial conditions. We pointed out the appearance 
of two distinct, non stationary, asymptotic regimes which occur depending 
on the temperature of the initial realization. 
For small values of $T_0$, similarities with the isothermal 
solution are found while for larger temperatures the density and velocity 
profiles are well reproduced also assuming a water-bag distribution. 

We propose a form of the density profile, with a cut-off in the tails, 
which fits well numerical data in all the explored regimes, allowing 
to derive a close analytical expression of the gravitational potential.
Moreover, a King-like velocity profile is shown to be in good agreement with
numerical data.  The asymptotic truncated profiles are thus a direct 
consequence of the evaporation from the finite box.

Finally we have also given numerical evidence of 
some scaling laws, which remain to be theoretically explained, but which are 
strongly related to the escaping process.

In the future, we plan to extend this study to the system of concentric 
spherical mass shells~\cite{henon} by introducing an external absorbing 
boundary in the configuration space as done here.

\vskip 1truecm
\noindent
{\bf Acknowledgements}
We thank an anonymous referee for his helpful comments, which led to
a substantial improvement of this paper. We acknowledge useful discussions
with E. Aurell, M.C. Firpo, M. Henon and A. Noullez. 
This work is part of the MURST-COFIN00 grant on {\it Chaos and localization 
in quantum and classical mechanics}. D.F. warmly acknowledges 
the research group {\it Dynamics of Complex Systems} in Florence  
for the kind hospitality.

\newpage

\end{document}